\documentclass[letterpaper,english,aps,prl,reprint,superscriptaddress]{revtex4-1}
\usepackage[T1]{fontenc}
\usepackage[latin9]{inputenc}
\usepackage{amsmath}
\usepackage{amssymb}
\usepackage{graphicx}
\usepackage{float}

\makeatletter

\pdfpageheight\paperheight
\pdfpagewidth\paperwidth

\providecommand{\tabularnewline}{\\}

\PassOptionsToPackage{position=top}{subfig}

\makeatother

\usepackage{babel}
\begin{document}
	\title{Quantum Hall states for Rydberg atoms with laser-assisted dipole-dipole interactions}
	\author{Tian-Hua Yang}
\thanks{These authors contribute equally to the work.}
	\affiliation{International Center for Quantum Materials and School of Physics, Peking University, Beijing 100871, China}
	\affiliation{Collaborative Innovation Center of Quantum Matter, Beijing 100871, China}
	\author{Bao-Zong Wang}
\thanks{These authors contribute equally to the work.}
	\affiliation{International Center for Quantum Materials and School of Physics, Peking University, Beijing 100871, China}
	\affiliation{Collaborative Innovation Center of Quantum Matter, Beijing 100871, China}
	\author{Xin-Chi Zhou}
	\affiliation{International Center for Quantum Materials and School of Physics, Peking University, Beijing 100871, China}
	\affiliation{Collaborative Innovation Center of Quantum Matter, Beijing 100871, China}
	\author{Xiong-Jun Liu}
	\thanks{Corresponding author: xiongjunliu@pku.edu.cn}
	\affiliation{International Center for Quantum Materials and School of Physics, Peking University, Beijing 100871, China}
	\affiliation{Collaborative Innovation Center of Quantum Matter, Beijing 100871, China}
	\affiliation{Institute for Quantum Science and Engineering and Department of Physics, Southern University of Science and Technology, Shenzhen 518055, China}	
	\affiliation{CAS Center for Excellence in Topological Quantum Computation, University of Chinese Academy of Sciences, Beijing 100190, China}

	\begin{abstract}
		Rydberg atoms with dipole-dipole interactions provide intriguing platforms to explore exotic quantum many-body physics. Here we propose a novel scheme with laser-assisted dipole-dipole interactions to realize synthetic magnetic field for Rydberg atoms in a two-dimensional array configuration, which gives rise to the exotic bosonic topological states. In the presence of an external effective Zeeman splitting gradient, the dipole-dipole interaction between neighboring Rydberg atoms along the gradient direction is suppressed, but can be assisted when Raman lights are applied to compensate the energy difference. With this scheme we generate a controllable uniform magnetic field for the complex spin-exchange coupling model, which can be mapped to hard core bosons coupling to an external synthetic magnetic field. The highly tunable flat Chern bands of the hard core bosons are then obtained and moreover, the bosonic fractional quantum Hall states can be achieved with experimental feasibility. This work opens an avenue for the realization of the highly-sought-after bosonic topological orders using Rydberg atoms.
	\end{abstract}
	
	\maketitle
	
	{\em Introduction.}--The two-dimensional (2D) electrons coupled to an external magnetic field in the perpendicular direction can fill into Landau levels, giving rise to the prominent quantum Hall (QH) effects~\cite{QHE1980,QHE1982}, whose discovery opened up the extensive search for topological states of quantum matter~\cite{Hasan2010,Qi2011,Yan2012,Chiu2016,Yan2017}. Unlike the electrons which are fermions, no quantum Hall states are obtained for noninteracting bosons coupled to external magnetic field, since the bosons are condensed to the ground state at zero temperature, rather than filling into an entire Landau band. To realize QH phase for bosons necessitates strong repulsive interactions, so that the Bose liquids become incompressible and the bosonic QH effects may be reached~\cite{bosonQHE1,bosonQHE2,bosonQHE3,bosonQHE4,bosonQHE5,bosonQHE6,rotateReview}. In comparison with fermionic counterparts, the bosonic integer and fractional QH states are all strongly correlated topological phases, being intrinsic~\cite{chiralbosonQHE,chiralbosonQHE1} or symmetry-protected topological orders~\cite{SPTbosonQHE,SPTbosonQHE1,SPTbosonQHE2}. Important attempts at achieving the QH regime have been made in bosonic systems like rotating Bose-Einstein condensates~\cite{rotateBEC}, Hostadter-Hubbard model~\cite{Hostadter}, and interacting photons~\cite{lightQHE}, while the feasibility of fully realizing 
	such strongly correlated topological phases in experiment is hitherto elusive. 
	
Recently, the exploration of novel correlated quantum states using Rydberg atoms attracted remarkable interests~\cite{review4}.~The Rydberg atoms can be arranged individually in array configuration through optical tweezers~\cite{Rydberg-tweezer3,Rydberg-tweezer4,Rydberg-tweezer5}. The highly excited internal states enable the long-range dipole-dipole interactions,
which generate effective hopping couplings between Rydberg atoms at different sites~\cite{Rydberg-lattice1,Rydberg-lattice2}. Such configuration simulates the hard-core bosons in lattice and provides versatile platforms to explore correlated bosonic quantum matter. Several important fundamental correlated phases have been observed in experiment, including quantum magnetism~\cite{Rydberg-spin1,Rydberg-spin2,Rydberg-spin4,Rydberg-spin5}, the 1D bosonic Su-Schrieffer-Heeger model~\cite{Rydberg-lattice2}, and 2D quantum spin liquid~\cite{Rydberg-QSL}. 
To further realize the bosonic QH phase with Rydberg arrays necessitates the generation of synthetic magnetic field which is associated with complex-valued dipole-dipole interactions.
The synthetic gauge fields are key ingredient to explore topological physics, and have been actively studied for ultracold atoms in optical lattices~\cite{Hostadter-Hamiltonian0,Hostadter-Hamiltonian1,Hostadter-Hamiltonian2,Hostadter-Hamiltonian3,LiuXJ2016NPJ,ChernBand1,ChernBand2,ChernBand3,ChernBand4,OpticalRamanLattice1,OpticalRamanLattice2,OpticalRamanLattice7,OpticalRamanLattice8,OpticalRamanLattice9}. Being intrinsically strongly correlated quantum simulators, the Rydberg arrays with synthetic magnetic fields are of great interests.
	
	In this letter, we propose a novel mechanism dubbed {\em laser-assisted dipole-dipole interaction} for realizing a tunable synthetic magnetic field for hard-core bosons simulated by Rydberg atoms in a 2D array configuration. The realized model is described by the Hamiltonian
	\begin{align}
		H= & \sum_{j_{x},j_{y}}(J_{x}b_{j_{x}+1,j_{y}}^{\dagger}b_{j_{x},j_{y}}+J_{y}e^{i\Phi j_{x}}b_{j_{x},j_{y}+1}^{\dagger}b_{j_{x},j_{y}}+h.c.)\nonumber \\
		& +\sum_{j_{x},j_{y}}(J_{d_{1}}e^{i\Phi j_{x}}b_{j_{x}+1,j_{y}+1}^{\dagger}b_{j_{x},j_{y}}+h.c.)\nonumber \\
		& +\sum_{j_{x},j_{y}}(J_{d_{2}}e^{i\Phi j_{x}}b_{j_{x}-1,j_{y}+1}^{\dagger}b_{j_{x},j_{y}}+h.c.),
		\label{Eq:Hamiltonian}
	\end{align}
	where $b_{i,j}^{\dagger}$ ($b_{i,j}$) creates (annihilates) a hard-core boson at site $(i,j)$ with particle number $\langle b_{i,j}^\dag b_{i,j}\rangle\leq1$, the coefficients $J_{x(y)}$ and $J_{d_{1(2)}}$ characterize the nearest neighbor (NN) hopping term along $x$ ($y$) direction and next-nearest neighbor (NNN) hopping terms in two diagonal directions, respectively. The hopping phase $\Phi$ represents a synthetic magnetic flux for the hard-core bosons, and is induced by the Raman laser-assisted dipole-dipole interactions. With the generated synthetic magnetic field, the flat Chern bands of the hard core bosons are obtained, with their flatness being drastically tuned by the diagonal $J_{d_1,d_2}$ terms. This study may pave the way for realizing bosonic QH states with Rydberg arrays. 
\begin{figure}[tp]
\includegraphics[width=1.0\columnwidth]{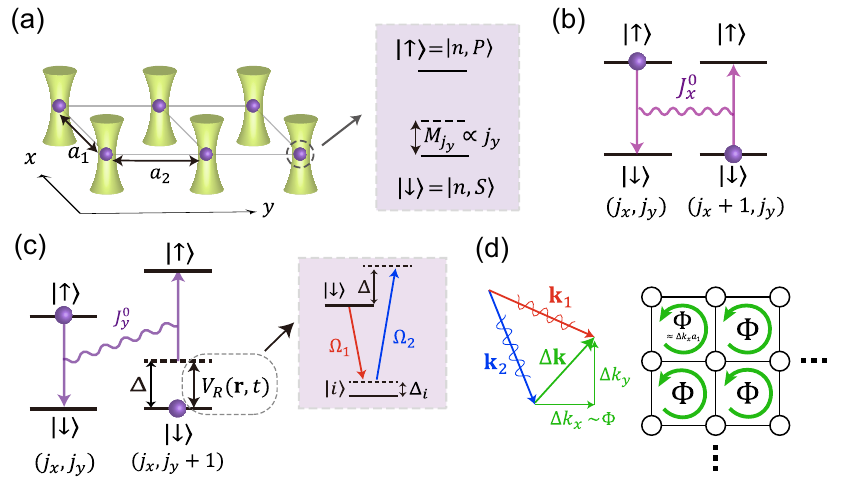}
\caption{\label{Fig1:setup} Sketch of the proposal. (a) Rydberg atoms trapped in optical tweezers to form a 2D array. An energy shift with gradient in $y$-direction modifies the energy difference between the two Rydberg states (effective Zeeman splitting for pseudospin). (b) Rydberg
			dipole-dipole interaction can induce spin exchange coupling between adjacent sites
			along the $x$-direction. (c) In $y$-direction, the spin exchange coupling
			is suppressed by the effective Zeeman energy offset. A two-photon Raman process compensates this energy offset and leads to
			the laser-assisted spin exchange couplings. (d) The laser-assisted exchange couplings have a spatially
			dependent phase, generating a synthetic magnetic flux in
			the 2D lattice.}
	\end{figure}
	
	{\em Laser-assisted dipole-dipole interactions.}--We consider the 2D rectangular array of Rydberg atoms, with lattice constants $a_{1,2}$ and each trapped in optical tweezers [see Fig.~\ref{Fig1:setup}(a)]. Two Rydberg states are chosen to simulate pseudo-spin-$1/2$ at each site, with $|\downarrow\rangle\equiv|n,S\rangle$ and $|\uparrow\rangle\equiv|n,P\rangle$. An effective Zeeman splitting $M_{j_y}$ between the two pseudospin states is introduced, with $M_{j_y+1}-M_{j_y}=\Delta$ along the $y$-direction, while the on-site energy along $x$-direction is uniform. The total Hamiltonian of the system
	\begin{equation}\label{totalHamiltonian}
		H=H_{\mathrm{dipole}}+H_{\mathrm{Zeeman}}+V_R(\bold r, t)
	\end{equation}
includes the bare dipole-dipole interactions~\cite{review4} which we take up to diagonal terms
\begin{align}
H_{\mathrm{dipole}}= &\sum_{j_{x},j_{y}}(J_{x}^{0}\sigma_{j_{x},j_{y}}^{+}\sigma_{j_{x}+1,j_{y}}^{-}+J_{y}^{0}\sigma_{j_{x},j_{y}}^{+}\sigma_{j_{x},j_{y}+1}^{-})\nonumber \\
 & +\sum_{j_{x},j_{y}}J_{d}^{0}\sigma_{j_{x},j_{y}}^{+}\sigma_{j_{x}+1,j_{y}\pm1}^{-}+h.c., \nonumber
 \label{eq:Hdip}
\end{align}
	the effective Zeeman energy gradient term
	\begin{equation}
		H_{\mathrm{Zeeman}}=\frac{1}{2}\sum_{j_{x},j_{y}}M_{j_{y}}\sigma_{j_{x},j_{y}}^{z}, \nonumber
		\label{eq:HZeeman}
	\end{equation}
	and the Raman coupling potential
	\begin{equation}
		V_R(\bold r, t)= \frac{\Omega_{1}\Omega_{2}^{\ast}}{\Delta_{i}} e^{i\Delta\mathbf{k}\cdot\mathbf{r}}e^{i(\omega_2-\omega_1) t} \sigma^x_{j_x,j_y} + h.c.. \nonumber
	\end{equation}
	In the above Hamiltonian, the dipole-dipole interaction leads to a spin exchange coupling $J_x^0$ between adjacent sites along the $x$-direction, as illustrated in Fig.~\ref{Fig1:setup}(b). The key ingredient of the scheme is that the bare exchange couplings $J_y^0$ and $J_d^0$ are suppressed by the relatively large Zeeman splitting offset $\Delta$, but can be further induced by applying the Raman coupling potential $V_R$ which is generated by two Raman lights with the Rabi-frequencies $\Omega_{1,2}$ and frequency difference $\omega_2-\omega_1 \approx \Delta$ such that the Zeeman energy offset $\Delta$ is compensated by the two-photon process. Specifically, this Raman process is obtained by coupling one of the pseudospins to an intermediate state $|i\rangle$ with detuning $\Delta_i$ [see Fig.~\ref{Fig1:setup}(c)]. With this configuration, the effective exchange couplings along the $y$ and diagonal directions are recovered by the Raman laser-assisted dipole-dipole interactions. Furthermore, the the wave vector difference
	$\Delta \bf{k} = \bf{k}_1 - \bf{k}_2$ of two Raman lights determines the phases of the induced exchange couplings which are responsible to the magnetic flux in the effective model [Fig.~\ref{Fig1:setup}(d)].
	
	\begin{figure*}[tp]
		\includegraphics[width=1.8\columnwidth]{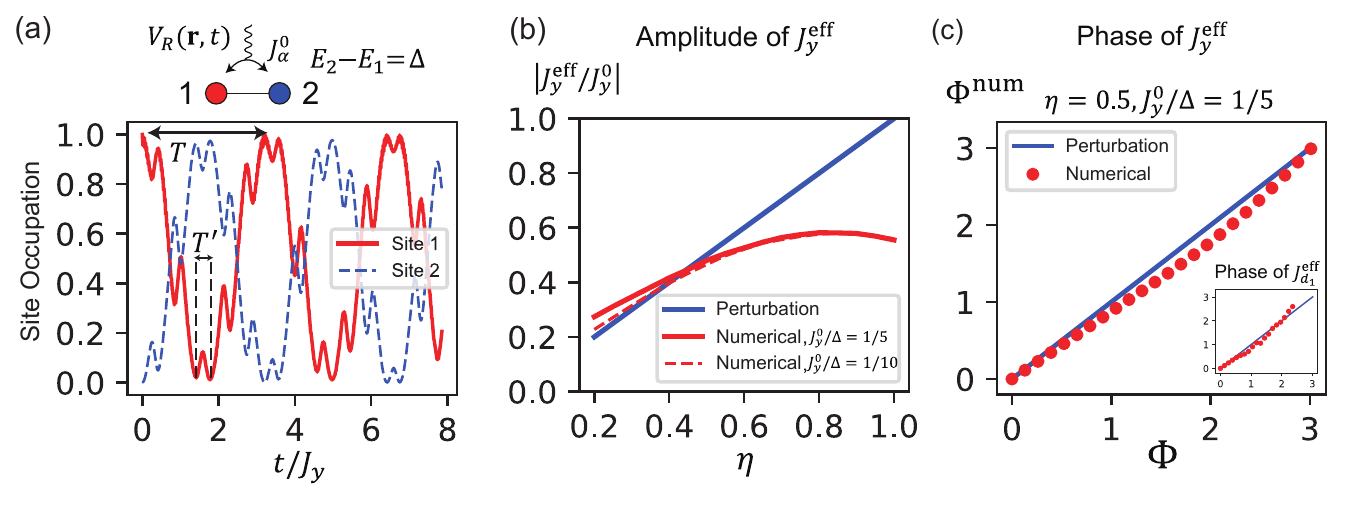} 
		\caption{\label{Fig2:evolution} Numerical simulation for the laser-assisted exchange couplings and the synthetic magnetic flux under $\phi_y=\pi$. (a) Raman coupling driven Rabi-oscillation on a two-site system, which has an energy offset $\Delta$ compensated by the Raman potential. A boson is initialized at site 1 and evolves afterwards. Take $\eta=2\left|\Omega_1\Omega_2/(\Delta\Delta_i)\right|=0.5$ and $\Delta=5J_y^0=10J_y$. The slow-oscillation is driven by the Raman coupling, while the fast oscillations of frequency $\Delta$ correspond to the off-resonant bare transition. (b) The amplitude of $J_y^{\mathrm{eff}}$ matches well with the perturbative result $\left|J_y^{\mathrm{eff}}\right|=\eta J_y^0$ when $\eta$ and $J_y^0/\Delta$ are small. When $\eta$ is large, $\left|J_y^{\mathrm{eff}}\right|$ deviates clearly from the perturbation results. (c) The phase of the exchange couplings $J_y$, $J_{d_1}$ and $J_{d_2}$ (not shown in figure). The numerical results show good coincidence with perturbation results.
}
	\end{figure*}
	With the above analysis we can compute the effective exchange couplings through a time-dependent perturbation theory (see Supplemental Material for details~\cite{supp})
\begin{equation}
  \begin{split}
		J_{y}^{{\rm eff}}= & J_{y}^{0}\frac{\Omega_{1}\Omega_{2}^{\ast}}{\Delta\Delta_{i}}e^{i(\Phi j_{x}+\phi_{y}j_{y})}(e^{i\phi_{y}}-1), \\
		J^{{\rm eff}}_{d_{1}(d_{2})}= & J_{d}^{0}\frac{\Omega_{1}\Omega_{2}^{\ast}}{\Delta\Delta_{i}}e^{i(\Phi j_{x}+\phi_{y}j_{y})}[e^{i(\phi_{y}+(-)\Phi)}-1],
\end{split}
\label{eq:JyEff}
\end{equation}
	where $\Phi=\Delta k_xa_1$ is a nontrivial phase generating flux in each plaquette, and the phase $\phi_y =\Delta k_ya_2$ tunes the strengths of $J_y$ and $J_{d_1(d_2)}$. The term $e^{i\phi_{y}j_{y}}$ is however trivial and can be gauged out. We then reach the effective spin model in a more compact form
	\begin{align}
		H_{\mathrm{eff}}= & \sum_{j_{x},j_{y}}(J_{x}\sigma_{j_{x}+1,j_{y}}^{+}\sigma_{j_{x},j_{y}}^{-}+J_{y}e^{i\Phi j_{x}}\sigma_{j_{x},j_{y}+1}^{+}\sigma_{j_{x},j_{y}}^{-}\nonumber \\
		& +h.c.)+\sum_{j_{x},j_{y}}(J_{d_{1}}e^{i\Phi j_{x}}\sigma_{j_{x}+1,j_{y}+1}^{+}\sigma_{j_{x},j_{y}}^{-}+h.c.)\nonumber \\
		& +\sum_{j_{x},j_{y}}(J_{d_{2}}e^{i\Phi j_{x}}\sigma_{j_{x}-1,j_{y}+1}^{+}\sigma_{j_{x},j_{y}}^{-}+h.c.),
		\label{Eq:Heff}
	\end{align}
where $J_y$ and $J_{d_1(d_2)}$ denote the amplitudes of the effective exchange couplings, and $J_x=J_x^0$. The above model is mapped to the Hamiltonian~\eqref{Eq:Hamiltonian} for hard-core bosons by defining the bosonic operator $b_{j}^{\dagger}=| \uparrow\rangle_{j}\langle\downarrow|_{j}$ for the pseudo-spin-$1/2$ at each site.
	
We note that the Eq.~\eqref{eq:JyEff} is obtained in the perturbative regime and holds precisely when $\Delta$ is large compared with the bare exchange couplings and the two-photon Raman coupling strength, namely $J_{y,d}^0/\Delta\ll1$ and $|\Omega_{1}\Omega_{2}^{\ast}|/(\Delta\Delta_{i})\ll1$. However, the generation of the magnetic flux through the laser-assisted dipole-dipole interactions is actually valid for more generic case beyond perturbative regime. The only difference is that for a moderate $\Delta$, higher-order processes and additional intermediate processes will also contribute to the effective exchange couplings, which mainly quantitatively modify the amplitudes in Eq.\eqref{eq:JyEff}, as we show below. 
	
We confirm the above results numerically by studying the hopping dynamics for a single boson along the $y$ direction or diagonal direction, as shown in Fig.~\ref{Fig2:evolution}.
We initialize the state of single boson occupying the site $1$, and numerically simulate the Rabi-oscillations by computing the dynamical evolution between the two sites from the original Hamiltonian~\eqref{totalHamiltonian}, with which we determine the numerical result of $J_y^{\rm eff}$ (the numerical study for $J_{d_1(d_2)}^{\rm eff}$ is similar, see Supplementary Material~\cite{supp}). 
Fig.~\ref{Fig2:evolution}(a) shows an example of the Rabi-oscillations, from which one can read off directly the amplitude of the $J_y^{\rm eff}$. From the phase accumulation in the wave function evolution, one can determine the phase $\varphi$ of the exchange coupling coefficient. Further, to obtain the numerical result of the magnetic flux per plaquette, denoted as $\Phi^{\rm num}$, we compute $\varphi(j_x)$ in two separate simulations for the two-site system along $y$ direction, respectively at $j_x=0$ and $j_x=1$. Then the flux is given by $\Phi^{\rm num}=\varphi(j_x=1)-\varphi(j_x=0)$~\cite{supp}. 
Based on this procedure and with different parameters, in Fig.\ref{Fig2:evolution} (b,c) we numerically obtain $\Phi^{(\rm num)}$ and $J_y^{(\rm eff)}$ (blue solid lines), and compare with the perturbation results in Eq.~\eqref{eq:JyEff} (red dashed lines).
We find that for relatively small $J_y^0/\Delta$ and $|\Omega_{1}\Omega_{2}^{\ast}|/(\Delta\Delta_{i})$, the numerical results of the amplitude of the laser-assisted exchange coupling $|J_y^{\rm eff}|$ match better those given from the perturbation theory [Fig.\ref{Fig2:evolution}(b)].
In comparison, the numerical results for the flux $\Phi^{\rm num}$ matches well the perturbation results in more generic results [Fig.\ref{Fig2:evolution}(c)]. With this we see that in the generic case the laser-assisted exchange couplings are induced, 
together with a nontrivial phase generating the magnetic flux in the effective model.
	
	Before proceeding we provide estimates for the model parameters in the real experiment. For the $^{87}$Rb atoms, for instance, we may take the primary quantum number $n\sim50$ for the Rydberg states, which are of the lifetime $\tau\sim100\mu\mathrm{s}$ at low temperature~\cite{Rydberglifetime}. The lattice constants $a_{1,2}$ can be taken to be $10\sim20\mu$m, for which the bare exchange coupling is about $J_{x,y}^0\approx1\sim2$MHz. Accordingly, it is sufficient to set the effective Zeeman splitting offset as $\Delta\approx5.0\sim10$MHz to suppress the bare exchange couplings along the $y$ and diagonal directions. When a Raman coupling with strength $\Omega_1\Omega_2/\Delta_i\sim0.25$ is applied, the effective coupling of magnitudes $J_y\sim0.6$MHz to $1.0$MHz is induced through numerical calculation.
	
	\begin{figure}[ht]
		\includegraphics[clip]{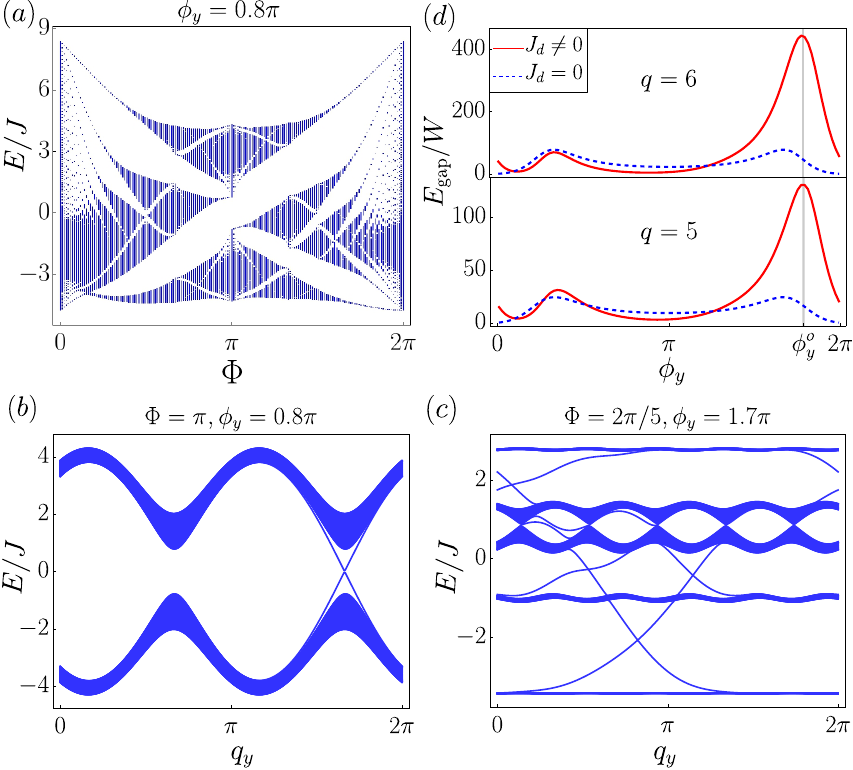}
		\caption{\label{Fig3:Hofstadter}The deformed Hofstadter butterfly
			and Quantum Hall bands modulated by the diagonal hoppings $J_{d_1(d_2)}$ when $J_x^0=J_y^0$. (a) The deformed Hofstadter butterfly is generated at $\phi_y=0.8\pi$. (b)-(c) The Chern bands for different $\phi_y$ and $\Phi$. (d) The flatness ratio $E_{\mathrm{gap}}/W$ versus $\phi_{y}$. The red solid lines show the flatness ratio for the realized model, in comparison with the case of setting $J_d=0$ by hand (blue dashed line corresponding). The maximal flatness is obtained at $\phi_y=\phi_y^o$.}
	\end{figure}
	As a key ingredient of the present scheme, the effective Zeeman splitting offset between neighboring sites can be realized with various approaches in the real experiment. 
	For example, one can apply additional optical lights, which can be set together with the optical tweezer lights, to couple one of the two Rydberg states say $|\downarrow\rangle$ and the ground state $5S$ for $^{87}$Rb atoms (or other low-energy normal states), giving an AC Stark shift to the Rydberg state $|\downarrow\rangle$. Using the same optical tweezer technique one can readily control the light field strength on each array at different $j_y$ sites to realize the required effective Zeeman splitting offset. Another direct approach is apply a magnetic field with spatial gradient along $y$ direction, which induces the real Zeeman energy splitting between the $S$ and $P$ Rydberg atoms. More details can be found in the Supplementary Material~\cite{supp}.
	
{\em Flat Chern bands for Rydberg states.}--We proceed to study the Chern band physics of the realized Hamiltonian~\eqref{Eq:Hamiltonian}, which exhibit novel features. In particular, in the presence of the NNN hopping $J_{d_{1(2)}}$, the energy spectra versus the flux $\Phi=(p/q)2\pi$ (with $p$ and $q$ being mutually prime integers) exhibits distinct characters in comparison with the conventional Hofstadter butterfly which is symmetric with respect to both $\Phi$ and energy~\cite{Harper(1955),Hofstadter(1976)}. Specifically, here the energy spectra are generically asymmetric, showing a deformed Hofstadter butterfly diagram [Fig.~\ref{Fig3:Hofstadter}(a)]. 
Interestingly, for the $\pi$-flux regime, the bulk is gapped with nonzero Chern number [Fig.~\ref{Fig3:Hofstadter}(b)], in contrast to the conventional case without diagonal terms, where the bulk is gapless~\cite{Hofstadter(1976)}. For $q=5$, a highly-flat lowest Chern band is obtained [Fig.~\ref{Fig3:Hofstadter}(c)].

The intriguing feature is that the NNN hopping coefficients $J_{d_{1(2)}}$ can drastically change the flatness ratio between the band gap $E_{\mathrm{\mathrm{gap}}}$ and band width $W$ regarding the lowest Chern band. 
Fig.~\ref{Fig3:Hofstadter}(d) shows numerically the flatness ratio (the red solid lines) versus $\phi_y$ which governs $J_y$ and $J_{d_1(d_2)}$ via Eq.~\eqref{eq:JyEff}, and for comparison the flatness ratio for the case of setting $J_d=J_{d_{1,2}}=0$ by hand is also given (the blue dashed lines). We find that the flatness of the lowest band is greatly improved in a large range of $\phi_y$. Especially, at $\phi_y=\phi_{y}^{o}\approx 1.7\pi$ for $J_x^0=J_y^0$, the diagonal hoppings $J_{d_{1}}=0.1e^{i0.55\pi}$ and $J_{d_{2}}=0.6e^{i0.85\pi}$, for which the flatness ratio is optimized to maximum and is very large. This feature enables a feasible way to realize bosonic fractional QH states.

\begin{figure}[t]
\includegraphics[width=0.9\columnwidth]{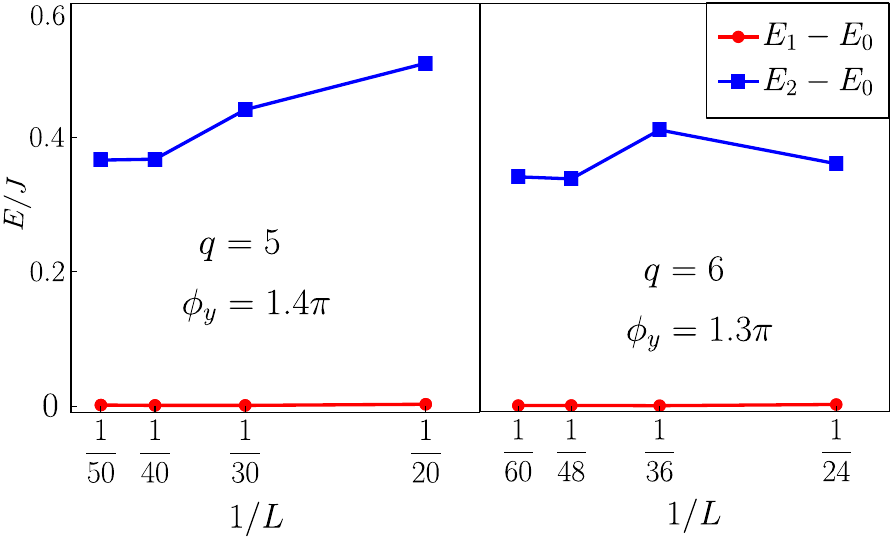}
\centering
		\caption{\label{Fig4:Low-energy-spectrum}$1/2$-fractional QH state shown from low energy spectra $E_{n}-E_{0}$
			versus the system size $1/L$, with $L=L_{x}\times L_{y}$. The results show the two-fold degeneracy of the many-body ground
			states, which have a finite gap separating from the excited states. Other parameters are taken $J_x=J_y=J$.}
\end{figure}
{\em Bosonic $1/2$ Laughlin state.}--The flat Chern bands for hard-core bosons facilitate the realization of bosonic fractional QH states. In comparison with rotating Bose-Einstein condensates~\cite{rotateReview}, the present Rydberg system realizes ideal Landau bands for hard-core bosons without necessitating fast-rotating condition. Also, unlike the Hofstadter model for ultracold atoms in optical lattice, the present model intrinsically reaches the strong interacting limit without suffering higher band effects. We denote the number of hard-core bosons as $N_{b}$, and the filling factor $\nu=N_b/N_{\varphi}$, where $N_{\varphi}$ is the total magnetic flux threading the 2D array. As a prominent example, we consider the filling $\nu=1/2$, the ground many-body wave function of this Bosonic Laughlin state reads~\cite{Laughlin(1987)}
	\begin{equation}
		\Psi_{gs}(z_{1},..,z_{N_{b}})=\prod_{j<k}(z_{j}-z_{k})^{2}\mathrm{exp}(-\sum_{i=1}^{N_{b}}|z_{i}|^{2}),
	\end{equation}
where $z_{j}=x_{j}+iy_{j}$ is the coordinate in the complex plane of the $j$th particle. The $\nu=1/2$ fractional QH state is characterized by two fundamental features. First, the many-body ground states have two-fold degeneracy. Second, the ground state manifold is separated from excitations with a finite gap. Below we confirm the two features based on exact diagonlization for a finite system of $L_{x}\times L_{y}$ sites with periodic boundary condition.
	
The numerical results are shown in Fig.~\ref{Fig4:Low-energy-spectrum}, where the hopping coefficients are set as $J_x=J_y=J$ for convenience at the phase $\phi_y=1.4\pi$ ($1.3\pi$) for $q=5$ ($6$). We compute the lowest three many-body eigenstates of the system, with energies $E_{0,1,2}$ and plot the spectra versus system size. We find the results are stabilized with sizes up to $5\times10$ for $q=5$ and $6\times10$ for $q=6$ with filling $\nu=1/2$. 
We see clearly that there is two-fold degeneracy for the many-body ground states as $E_{1}-E_{0}\approx0$, while the excitation gap $E_{\rm gap}=E_{2}-E_{1}$ approaches an appreciable magnitude at large-size limit. This yields the	gap $E_{\rm gap} = 0.37J$ for $q=5$ and $E_{\rm gap} = 0.35J$ for $q=6$ for the present fractional QH phase.
	
{\em Conclusions}.--We have proposed a novel scheme dubbed {\em laser-assisted dipole-dipole interactions} for Rydberg atoms to realize synthetic magnetic field and 2D bosonic QH states. The dipole-exchange interaction along one direction of the 2D Rydberg array is suppressed by setting an effective Zeeman splitting gradient, but can be assisted by applying a two-photon Raman coupling process which compensates the neighboring-site Zeeman energy offset and generates nontrivial gauge flux for the spin-exchange model. The tunable flat Chern bands of hard-core bosons and the bosonic fractional QH states can be obtained feasibly, with the $1/2$-Laughlin state being illustrated. This work introduces a basic scheme of laser-assisted dipole-dipole interaction which can greatly expand the capability of engineering Rydberg atoms coupling to synthetic gauge fields and can be broadly applied to various Rydberg array configurations, 
hence may open an avenue to realize exotic correlated topological models and explore the highly-sought-after bosonic topological orders with experimental feasibility. 

{\em Acknowledgement}.--We thank Shi Yu and Zheng-Xin Liu for fruitful discussions. This work was supported by National Key Research and Development Program of China (2021YFA1400900), the National Natural Science Foundation of China (Grants No.11825401, No.12104205), 
and the Strategic Priority Research Program of Chinese Academy of Science (Grant No. XDB28000000).

	\onecolumngrid
	
	\renewcommand{\thesection}{S-\arabic{section}}
	\setcounter{section}{0}  
	\renewcommand{\theequation}{S\arabic{equation}}
	\setcounter{equation}{0}  
	\renewcommand{\thefigure}{S\arabic{figure}}
	\setcounter{figure}{0}  
	\newcommand{\diff}{\mathrm{d}}
	\newcommand{\biaoti}{\fontsize{12pt}{\baselineskip}\selectfont}

	\indent

	\begin{center}
\large \textbf{\large Supplementary Material:\\Quantum Hall states for Rydberg atoms with laser-assisted dipole-dipole interactions}
\end{center}

	\section{Time-dependent Perturbation Theory}
	
	In this section, we derive the effective exchange couplings $J_y^{\rm eff}$ and $J^{\rm eff}_{d_1(d_2)}$ using the time-dependent perturbation theory. Start with the Hamiltonian $H=H_{\rm{dipole}}+H_{\rm{Zeeman}}+ V_{R}(\bold r, t)$, with $H_{\rm{dipole}}$ and $H_{\rm{Zeeman}}$ given in main text. The Raman process, which consists of two Raman lasers coupling $|\downarrow\rangle$ to a low-lying intermediate state $|i\rangle$, can be described as
	\begin{equation}
		V_{R}(\bold r, t) = \sum_j \bigr[\left(\Omega_1e^{i\mathbf k_1\cdot\mathbf r}e^{-i\omega_1 t} + \Omega_2 e^{i\mathbf k_2\cdot\mathbf r}e^{-i\omega_2 t}\right)|\downarrow\rangle_{j_x,j_y}\langle i|_{j_x,j_y}+h.c.\bigr]+(\Delta_i-\omega_1)|i\rangle_{j_x,j_y}\langle i|_{j_x,j_y}.
	\end{equation}

\begin{figure}[h]
		\includegraphics[width=0.8\columnwidth]{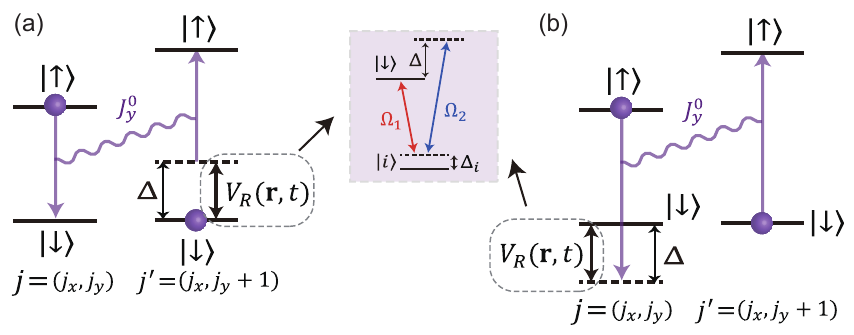}
		\caption{\label{Fig:Supp} The possible laser-assisted transition processes between two sites. In both cases, a two-photon process between $|\downarrow\rangle$ and $|i\rangle$ on one site compensates an energy difference $\Delta$. The intermediate state can be (a) $|j=\downarrow,j^\prime=i\rangle$ or (b) $|j=i,j^\prime=\downarrow\rangle$ . The two processes add up to give the total effective transition moment. }
	\end{figure}

	Consider the assisted hopping $J_y^\mathrm{eff}$ between two adjacent sites $j=(j_x,j_y)$ and $j^\prime=(j_x,j_y+1)$. We denote $|a\rangle=|j=\uparrow,j^\prime=\downarrow\rangle$ and $|b\rangle=|j=\downarrow,j^\prime=\uparrow\rangle$.  The exchange couplings between  $|a\rangle$ and $|b\rangle$ are suppressed by the large Zeeman splitting offset $\Delta$, and are further recovered by appling the Raman coupling potential $V_R$: a two-photon process compensating the energy offset can take place either on the site $j$ or $j^\prime$, as shown in the Fig.~\ref{Fig:Supp}.
	In the first case, the compensation is essentially obtained by the Raman coupling between spin down state $|\downarrow\rangle_{j}$ and intermediate state $|i\rangle_{j}$. Therefore, all information about this channel is contained in the three-dimensional subspace spanned by $|a\rangle,|b\rangle$ and $|c\rangle=|j=i,j^\prime=\uparrow\rangle$. Similarly, the compensation on site $j^\prime$ happens in the subspace spanend by $|a\rangle,|b\rangle$ and $|c^\prime\rangle=|j=\uparrow,j^\prime=i\rangle$. Therefore, we may calculate the assisted hopping amplitude in such three-level subspaces.
	
	Consider the channel where compensation happens at site $j^\prime$. The effective Hamiltonian under the basis $(|a\rangle,|c^\prime\rangle,|b\rangle)$ can be written as
	\begin{equation}
		H_{\mathrm{TL}}=\left(\begin{array}{ccc}
			E_{a} & \Omega_1e^{i\mathbf k_1\cdot\mathbf r^\prime}e^{-i\omega_1 t} + \Omega_2 e^{i\mathbf k_2\cdot\mathbf r^\prime}e^{-i\omega_2 t} & J_y^0\\
			\Omega_1^\ast e^{-i\mathbf k_1\cdot\mathbf r^\prime}e^{i\omega_1 t} + \Omega_2^\ast  e^{-i\mathbf k_2\cdot\mathbf r^\prime}e^{i\omega_2 t} & E_{c^\prime}\\
			J_y^0 &  & E_{b}
		\end{array}\right). \label{eq:Ham}
	\end{equation}
	The diagonal terms are the energies for the three states, respectively, satisfying $E_b-E_a=\Delta$ and $E_{c^\prime}-E_a=\Delta_i-\omega_1$. For simplicity of notation, we denote $\Omega_{l}(\mathbf r)=\Omega_le^{i\mathbf k_l\cdot\mathbf r}$ for $l=1,2$. Transferring to interaction picture gives
	\begin{equation}
		H_{\mathrm{TL}}^{(\mathrm{i})}=\left(\begin{array}{ccc}
			0 & \Omega_{1}(j^\prime)e^{i(-\omega_{1}-E_{c^\prime}+E_{a})t}+\Omega_{2}(j^\prime)e^{i(-\omega_{2}-E_{c^\prime}+E_{a})t} & J_y^0e^{-i\Delta t}\\
			\Omega_{1}(j^\prime)^{\ast}e^{-i(-\omega_{1}-E_{c^\prime}+E_{a})t}+\Omega_{2}(j^\prime)^{\ast}e^{-i(-\omega_{2}-E_{c^\prime}+E_{a})t} & 0\\
			J_y^0e^{i\Delta t} &  & 0
		\end{array}\right).
	\end{equation}
	Employing time-dependent pertubation theory, we consider possible channels for the transition $|b\rangle\langle a|$ in the Dyson series. The first-order process is the bare transition, which
	is off-resonant. There are no second-order processes for this transitions. Two third-order
	processes exist, manifesting themselves in the Dyson series as
	\begin{equation*}
		U^{(3)}(t)=-i^{3}\int_{0}^{t}\mathrm{d}t_{1}\int_{0}^{t_{1}}\mathrm{d}t_{2}\int_{0}^{t_{2}}\mathrm{d}t_{3}H_{31}(t_{1})H_{12}(t_{2})H_{21}(t_{3})+H_{31}(t_{1})H_{13}(t_{2})H_{31}(t_{3})\label{eq:U3t}.
	\end{equation*}
	Recursive integration yields multiple terms, but most are off-resonant
	and suppressed. Notably, a term in the first integrand reads
	\begin{equation}
		U_{\mathrm{Raman}}^{(3)}=\frac{\Omega_{1}(j^\prime)\Omega_{2}(j^\prime)^{\ast}J_y^0\left(e^{-i(\Delta+\omega_{2}-\omega_{1})t}-1\right)}{(-\omega_{1}-E_{c^\prime}+E_{a})(\omega_{2}-\omega_{1})(\Delta+\omega_{2}-\omega_{1})}.
	\end{equation}
	This is an equivalent Rabi transition of amplitude $\frac{\Omega_{1}(j^\prime)\Omega_{2}(j^\prime)^{\ast}J_y^0}{\Delta_{i}\Delta}$
	and detuning $\Delta-(\omega_{1}-\omega_{2})$, which is resonant when $\omega_1-\omega_2=\Delta$. A similar derivation goes when we choose $|c\rangle$ as the intermediate state. Combining these two channels, we have a resonant transition amplitude
	\begin{equation}
		J_y^{\mathrm{eff}}=\frac{\Omega_1\Omega_2^\ast}{\Delta\Delta_i}e^{i(\Delta k_xa_1j_x+\Delta k_ya_2j_y)}\left(e^{i\Delta k_ya_2}-1\right)J_y^0.
	\end{equation}
	Similarly one can derive the expression for $J^{\rm eff}_{d_1(d_2)}$. This leads to Eq.~(3) of the main text.
	
	Processes of other order in the perturbation series may lead to deviations from the above result. Some higher-order processes may affect the effective hopping,
	because a resonance can also be recovered by multiple compensations.
	Each two-photon process corrects $J_y$ by multiplying a factor $\eta=\frac{2|\Omega_{1}\Omega_{2}|}{|\Delta_{i}\Delta|}$ (a factor $2$ appears here because each compensation process can happen on either of the two sites), so resonant $2n$-photon processes have an amplitude of around $\eta^nJ_y^0$. When $\eta$ is finite such that higher powers of $\eta$ cannot be ignored, these processes will also contribute to $J_y^{\mathrm{eff}}$. This at most modifies the amplitude of $J_y^{\mathrm{eff}}$ but not its phase, however, since the condition of resonance makes additional phase factors cancel. On the other hand, when $J_y^0/\Delta$ is finite, the off-resonant bare transition (zeroth-order) is not fully suppressed. Naively, this increases the amplitude of the transition and dilute the phase, since the zeroth-order process has no phase. However, the effect of this bare transition may not manifest itself as a simple modification of $J_y^{\mathrm{eff}}$, since a resonant process and an off-resonant one cannot be added straightforwardly. In experiments, the exact phase and amplitude of $J_y^{\mathrm{eff}}$ may be determined through a calibration process by sweeping the parametric space.
	
	\section{Experimental Parameters}
	
	\begin{table*}[tp]
		\begin{tabular}{|c|c|c|c|}
			\hline
			Quantity & Typical Value & Quantity & Typical Value\tabularnewline
			\hline
			\hline
			Rydberg principal quantum number $n$ & $50$ & Rydberg state lifetime & around $100\mu\mathrm{s}$\tabularnewline
			\hline
			Lattice spacing (in $x$-direction) $a_1$ & $15\mu\mathrm{m}$ & Raman laser wavelength & $1000\mathrm{nm}$\tabularnewline
			\hline
			Rydberg atomic radius \cite{low2012anexperimental} & $0.16\mu\mathrm{m}$ & Detuning of Raman processes $\text{\ensuremath{\Delta_{i}}}$ & $25\mathrm{GHz}$\tabularnewline
			\hline
			Dipole-dipole interaction strength $J$ & $1\mathrm{MHz}$ & Energy difference between 5S and 5P & around $90\mathrm{THz}$\tabularnewline
			\hline
			Detuning $\Delta$ & $10\mathrm{MHz}$ & $\Omega_{1,2}$ & $250\mathrm{MHz}$\tabularnewline
			\hline
			Fine structure of Rydberh states & around $1\mathrm{GHz}$ & Transition dipole moment $\langle50,S|er|6,P\rangle$ & $0.01ea_{0}$\tabularnewline
			\hline
			Hyperfine splitting of Rydberg states & less than $200\mathrm{kHz}$ & Raman laser field strength & $3\times10^{5}\mathrm{V}/\mathrm{m}$\tabularnewline
			\hline
			Spacing between $n$ and $n+1$ Rydberg states & $60\mathrm{GHz}$ & The minimal Raman laser angle & $5^{\circ}$\tabularnewline
			\hline
		\end{tabular}
		
		\caption{Estimates of values for the relevant physical quantities.}\label{tableS1}
	\end{table*}
	
	In this section, we give the estimate of the orders of magnitude of the relevant experimental parameters. To be specific, the data given here are based on ${}^{87}$Rb atoms. We choose Rydberg states $|\downarrow\rangle=|50,S\rangle$ and $|\uparrow\rangle=|50,P\rangle$ as the pseudospin. A dipole-dipole interaction of strength $J_{ij}=\frac{C_3}{r_{ij}^3}$ exists between such two states, where $C_3\approx 3400\mathrm{MHz}\cdot\mu\mathrm{m}^{3}$ \cite{barredo2015coherent,review5}. On a rectangular lattice with lattice constants $a_{1,2}$, we have
\begin{equation}
J_x^0=\frac{C_3}{a_1^3}, \ J_y^0=\frac{C_3}{a_2^3},
\end{equation}
and
\begin{equation}
J_{d_1}^0=J_{d_2}^0=\frac{C_3}{(a_1^2+a_2^2)^{3/2}}.
\end{equation}
Since $J_y^{\mathrm{eff}}$ has an amplitude smaller than $J_y^0$, we may choose $a_2$ slightly smaller than $a_1$ to make $J_y^{\mathrm{eff}}=J_x^0$. For example, with $\phi_y=\pi$ and $\frac{\Omega_1\Omega_2}{\Delta\Delta_i}=\frac{1}{4}$, we can choose $a_1=15\mu\mathrm{m}$ and $a_2=2^{-\frac{1}{3}}a_1\approx 12\mu\mathrm{m}$, so that $J_x^0=J_y^{\mathrm{eff}}=J=1\mathrm{MHz}$. In this case $J_{d_1}^{\mathrm{eff}}=J_{d_2}^{\mathrm{eff}}=0.24\mathrm{MHz}$ when $\Phi=0$. The next resonant term is the next neartest neighbor hopping in $x$-direction, with an amplitude $J_{x^{(2)}}^0=J_x^0/2^3=0.125\mathrm{MHz}$, which we ignore. Van der Waals interaction can also be ignored, as it is well less than $100\mathrm{kHz}$ in this case~\cite{beguin2013directmeasurement}. Given the interaction strength, the effective Zemman energy gradient $\Delta$ can be chosen as $10\rm{MHz}$, giving a ratio $\frac{J_y^0}{\Delta}=\frac{1}{5}$.
	
	We have to ensure that no other undesired states or processes mix into our Hamiltonian. The two pseudospin states are separated by an energy of about $20\rm{GHz}$.	The fine structure splittings of $n=50$ Rydberg states are about hundreds of $\rm{MHz}$ \cite[Chap.~16]{gallagher1994rydberg} and the hyperfine structure splittings are at the level of $200\rm{kHz}$~\cite{ramos2019measurement}. We can see that the effective Zeeman splitting is much larger than the hyperfine energy and small enough compared to fine structure splittings. The intermediate state $|i\rangle$ is chosen as one of the 6P states. The Raman lasers would only couple $|\downarrow\rangle$ to this intermediate state, but not $|\uparrow\rangle$, because a single-photon process reverses parity and the $|50,P\rangle (|\uparrow\rangle) \to |6,P\rangle$ transition is forbidden. Another process $|50,P\rangle (|\uparrow\rangle) \to |6,S\rangle$ is possible from the point of view of parity, but is suppressed by an energy detuning of about $90\mathrm{THz}$. Other possible processes are detuned even more. Thus, as long as one chooses $\Delta_i<100\mathrm{GHz}$, other processes are at least three orders of magnitude smaller.
	
The strengths of the Raman lasers are labeled as $\Omega_{1,2}$, which equals to the electric field strength of the laser times the transition dipole moment. We can estimate that the transition dipole moment $\langle50,S|er|6,P\rangle\approx 0.01ea_{0}$, where $a_{0}$ is the Bohr radius \citep{low2012anexperimental}. This means that $\Omega_{1,2}(\mathrm{Hz})=803E_{1,2}(\mathrm{V}/\mathrm{m})$,with $E_{1,2}$ being the electric field strength. Thus experimental lasers can reach levels where $\Omega_{1,2}\sim$ hundreds of $\mathrm{MHz}$. An exemplary data would be $\Delta_i=25\mathrm{GHz}$ and $\Omega=250\mathrm{MHz}$. The Raman lasers would have an approximate wavelength of $\lambda =1000\mathrm{nm}$ as they couple 50S to 6P~\citep{lampen2018longlived}.The flux $\Phi=\Delta k_{x} a_1$ and the phase $\phi_y=\Delta k_{y} a_1$ enerated by Raman coupling are expected to be on the order of $\pi$. Given the lattice constants chosen above, $|\mathbf{k}|\approx\frac{1}{20}|\frac{2\pi}{\lambda}|$
	so the two lasers should have an intersecting angle of around $5^{\circ}$. To
	make the flux be accurate at or over $0.1\pi$ level, one would require tuning
	the laser angles at $0.5^{\circ}$ level or better, which is well achievable in experiment.
	
	To observe the correlated effects, the lifetime of the system should large compared with characteristic time defined by the inverse of the systems's energy scale. For $J=1\mathrm{MHz}$, a lifetime of around $100\mu\mathrm{s}$ is desirable. A Rydberg state of principal quantum number $n\approx 50$ can indeed have a lifetime of over $100\mu\rm{s}$ at low temperatures~\cite{beterov2009quasiclassical}. By coupling a Rydberg state to a 6P state, decay from the 6P state will also affect the lifetime. The lifetime of such processes should be much larger than that of the Rydberg state. The 6P states on their own have a lifetime of  around $100\mathrm{ns}$ \citep{gomez2004lifetime}. Under a detuned coupling, the wave function on that state would be $\frac{2\Omega_{1,2}}{\Delta_{i}}$, so the lifetime will be prolonged by a factor $(2\Omega_{1,2}/\Delta_{i})^{-2}$. Choosing $\frac{\Omega_{1,2}}{\Delta_{i}}<\frac{1}{100}$ would be sufficient to ensure
	a lifetime much larger than $100\mu\mathrm{s}$.
	
	For the creation of the effective Zeeman splitting using AC Stark effect,
	we can introduce an additional two-photon effective Raman coupling together with each optical tweezer, coupling the pseudospin state $|\downarrow\rangle=|50,S\rangle$ to the ground state $|5,S\rangle$. With a similar parity argument, $|\uparrow\rangle$ will not be coupled. This produces an energy shift $\Omega_{\mathrm{Stark}}^2/\Delta_{\mathrm{Stark}}$. With $\Omega_{\mathrm{Stark}}=200\mathrm{MHz}$ and $\Delta_{\mathrm{Stark}}=1\mathrm{GHz}$, a detuning of several tens of $\mathrm{MHz}$ can be achieved. By letting $\Omega_{\mathrm{Stark}}$ vary from $0$ to $200\mathrm{MHz}$ along $y$ direction, it is sufficient to realize the required effective Zeeman splitting gradient. The table~\ref{tableS1} shows typical parameter conditions.

	\section{Numerical simulation of two-site dynamics}
	
	\begin{figure}[tp]
		\includegraphics[width=1.0\columnwidth]{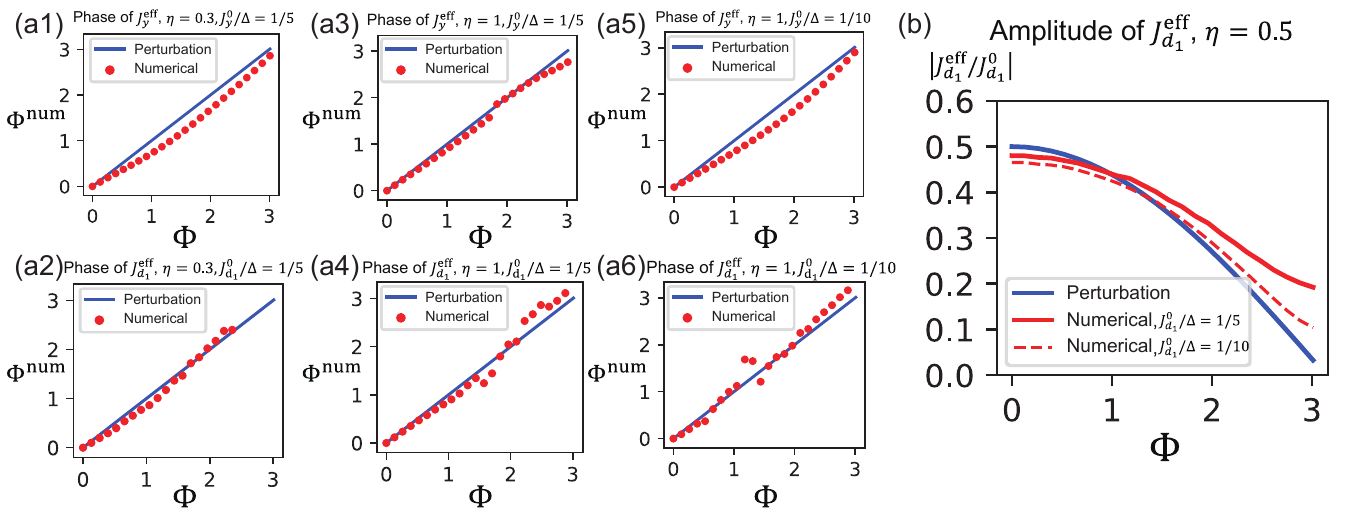} 
		\caption{\label{FigS2:evolution} (a1-a6) The site-dependent phase of $J_y^{\mathrm{eff}}$ and $J_{d_1}^{\mathrm{eff}}$ are plotted versus the applied $\Phi$ of the Raman potential under several sets of parameters, and with $\phi_y=\pi$. Aside from small numerical errors, the result matches the expectation that $J_\alpha^{\mathrm{eff}}\propto e^{i\Phi j_x}$ for $\alpha=y, d_1$. This shows that the generated flux is robust under a wide range of $\eta=\left|\frac{2\Omega_1\Omega_2}{\Delta\Delta_i}\right|$ and $J_y^0/\Delta$, even in regions where the amplitude of the effectie transition deviates significantly from the perturbative result. (b) The amplitude of $J_{d_1}$ roughly matches the perturbative formula $\left|J_{d_1}^{\mathrm{eff}}\right|=\eta J_{d_1}^0\sin\left(\frac{\phi_y+\Phi}{2}\right)$. The match is better at small $J_{d_1}^0/\Delta$. Both results for $J_{d_1}$ also hold for $J_{d_2}$.}
	\end{figure}
Suppose there are two sites with a resonant transition $J_{\alpha} e^{i\varphi}$
between them, where $J_{\alpha}$ is a real number represents the strength of transition. The Hamiltonian in two-dimensional subspace reads
$H_{2}=J_{\alpha}\left(\cos\varphi\sigma_{x}+\sin\varphi\sigma_{y}\right)$.
Then the time evolution operator would be
\begin{align}
e^{-iH_{2}t}=\left(\begin{array}{cc}
\cos\left( J_{\alpha} t \right) & ie^{-i\varphi}\sin \left(J_{\alpha}t \right)\\
-ie^{i\varphi}\sin \left( J_{\alpha} t \right) & \cos \left( J_{\alpha}t  \right)
\end{array}\right) .
\end{align}
When a particle is initially placed at the site 1, the particle wave function would evolute with the time as
\begin{align}
    |\psi(t)\rangle\propto\left(\cos\left(J_{\alpha}t\right),-ie^{i\varphi}\sin\left(J_{\alpha}t\right) \right)^{T}.
\end{align}
Therefore, the phase difference between two sites is obtained as $\varphi-\frac{\pi}{2}$ and $J_{\alpha}$ can be read off from the oscillation frequency of time evolution.

We take the simulation on the two sites with an energy difference $\Delta$ and Raman
compensations on both sites. The Hamiltonian reads
\begin{align}
H_{2\rm{s}}= &{}\Delta|2\rangle\langle2|+J_{\alpha}^{0}\left(|2\rangle\langle1|+|1\rangle\langle2|\right) \nonumber  \\
&{}+\left(\Omega_{1}(j_{1})e^{-i\omega_{1}t}+\Omega_{2}(j_{1})e^{-i\omega_{2}t}\right)\left(|i\rangle\langle1|+|1\rangle\langle i|\right) \nonumber \\
&{}+\left(\Omega_{1}(j_{2})e^{-i\omega_{1}t}+\Omega_{2}(j_{2})e^{-i\omega_{2}t}\right)\left(|i\rangle\langle2|+|2\rangle\langle i|\right),
\end{align}
where $|i\rangle$ is the intermediate state of Raman coupling,  $j_{1}=(j_{x},j_{y})$ and $j_{2}=(j_{x}+\delta j_{x},j_{y}+\delta j_{y})$ are two adjacent sites.
From the main text we know that the effective hopping is
\begin{equation}
J_{\alpha}^{\mathrm{eff}}=\frac{\Omega_{1}\Omega_{2}}{\Delta\Delta_{i}}J_{\alpha}^{0}\left(e^{i\left(\Phi\delta j_{x}+\phi_{y}\delta j_{y}\right)}-1\right)e^{i\left(\Phi j_{x}+\phi_{y}j_{y}\right)}.
\end{equation}
Therefore, we have
\begin{align}
J_{\alpha} &=2\frac{\Omega_{1}\Omega_{2}}{\Delta\Delta_{i}}\sin\left(\frac{\Phi\delta j_{x}+\phi_{y}\delta j_{y}}{2}\right)J_{\alpha}^{0} , \\
\varphi &=\Phi j_{x}+\phi_{y}j_{y}+\frac{\Phi\delta j_{x}+\phi_{y}\delta j_{y}+\pi}{2}  .
\end{align}
Note that the flux $ \Phi = \varphi(j_{x}+1,j_{y})-\varphi(j_{x})$. We perform the real-time numerical simulations for the dynamical evolutions from the two-site Hamiltonian and fit the parameters to confirm
the amplitudes and phases of $J_{y}$, $J_{d_{1}}$ and $J_{d_{2}}$.
In the simulation of $J_{y}$, we set for convenience that $\delta j_{x}=0$
and $\delta j_{y}=1$, and $j_{x}=j_{y}=0$ in light of the translational symmetry.
Then we take the same numerical simulation at $j_{x}=1$ and $j_{y}=0$, and take the difference of $\varphi$
measured in the two either cases to confirm the induced flux and that $J_{y}\propto e^{i\Phi j_{x}}$.
The absolute value of the phase $\varphi$, which is not important, may be complicate and influenced by multiple factors, however the phase difference (flux) is fairly stable (main text).
Similar simulation for $J_{d_{1}}$ and $J_{d_{2}}$ have been performed by taking
$\delta j_{x}=\pm1$ and $\delta j_{y}=+1$ [see Fig.~\ref{FigS2:evolution}].

\end{document}